\documentstyle[aps,prd,eqsecnum]{revtex}

\newcommand{\bmit}[1]{\mbox{\boldmath $#1$}}
\newcommand{\xin}{\bmit{x_{(i)}}}
\newcommand{\xo}{\bmit{x_{(o)}}}
\newcommand{\mxi}{|\bmit{x_{(i)}}|}
\newcommand{\mxo}{|\bmit{x_{(o)}}|}
\newcommand{\oti}{\tilde{\omega}_i}
\newcommand{\otio}{\tilde{\omega}_{i0}}
\newcommand{\ote}{\tilde{\omega}_o}
\newcommand{\oteo}{\tilde{\omega}_{o0}}

\newcommand{\ca}{c_{(\lambda)\aleph}}
\newcommand{\da}{d_{(\lambda)\aleph}}
\newcommand{\dio}[1]{d_{(io)#1}}

\newcommand{\half}{\frac{1}{2}}
\newcommand{\Ai}[2]{A_{(i)#1}^{\ \ \ \ #2}}
\newcommand{\Ao}[2]{A_{(o)#1}^{\ \ \ \ #2}}

\begin{document}

\draft
\title{Gravitational Radiation from Triple Star Systems}
\author{Enrico Montanari\footnote{Electronic address: 
montanari@fe.infn.it}, Mirco Calura, and Pierluigi Fortini}
\address{Department of Physics, University of Ferrara and
INFN Sezione di Ferrara, Via Paradiso 12,
I-44100 Ferrara, Italy}
\maketitle

\begin{abstract}
We have studied the main features of the gravitational 
radiation generated by an astrophysical system constituted
of three compact objects attracting one another 
(only via gravitational interaction) in such 
a manner that stable orbits do exist. We have limited our analysis to 
systems that can be treated with perturbative methods.
We show the profile of the gravitational waves 
emitted by such systems. 
These results can be useful within the framework of the new 
gravitational astronomy 
which will be made feasible by means of the new 
generation of gravitational detectors such as LISA in a no longer 
far future.
\end{abstract}

\section{Introduction}

A great deal of efforts are aimed at the detection of 
gravitational waves. Several large research programmes are 
under development all over the world~\cite{fra,pis}. Consequently the 
theoretical study of gravitational radiation generated by 
astrophysical systems is quite a relevant topic. So far this study has 
been mainly focused on such sources as supernovae explosions, pulsars 
and compact binaries. In particular these latter sources have been 
widely investigated throughout all the stages of their 
evolution~\cite{ll2,peters,wise,zhuge,moreno}. 

Space based interferometric detectors such as LISA are sensitive to 
gravitational waves in the frequency band 
$10^{-4}$--$10^{-1}\,Hz$~\cite{hough,cutler}. 
Many known gravitational waves sources, such as short--period binary 
star systems, are contained in this bandwidth~\cite{lipunov,hils}. 
However their waveforms could be biased by spurious effects such as 
those due to 
tidal forces, spinning of the stars, post--Newtonian corrections, 
presence of interstellar medium  
or of a third body. In order to recognize a gravitational wave from 
the experimental data, the expected wave form of the signal must be 
known in a highly accurate way. It is for this reason 
that many results have been obtained in order to take into account 
some of the above 
effects~\cite{bla96,khr96,wil96,bla96a,ruf96,apo96}. 
However, to our knowledge, a 
solution to the problem of the waveform from triple star system is 
still lacking. 
In this paper we study the characteristics 
of the gravitational radiation produced by such a source, emphasizing 
the substantial differences with the waves originating from a simple 
binary system, providing a tool in order to recognize 
astrophysical systems from the analysis of the gravitational 
signal detected.
Future gravitational astronomy will allow to open a 
new window on the universe besides the usual way to investigate the 
sky with electromagnetic devices or the relatively new neutrino 
astronomy (whose birth could be the 
``neutrino description'' of the supernova explosion 
1987A~\cite{hir87,hir88,bionta,bratton}). 
In fact gravitational radiation detection could 
be the only way to discover and study all that astrophysical systems 
which are not visible with standard methods because they do not  
emit electromagnetic radiation nor neutrinos. In particular, 
the existence of a black--hole could be argued by its effect on the 
gravitational radiation emitted by a binary system. 
In this paper we show that the gravitational signal of a 
binary system is modulated by a perturbing third body. We also show 
that the signal is different in form. Besides, triple 
star systems formed by a close binary and a third body radiate the 
same amount of energy of a close binary (or even more) in the region 
of interest for LISA. This means that this detector can have the same 
sensitivity for both kinds of system provided that the waveform of the 
radiation from triple star systems is known. In fact, with this 
knowledge, one can analyze the two time series that will be the data 
of LISA, searching for periodic sources not having the typical form of 
the signal from a binary system~\cite{cutler}.
This can be done without any change in the detector project, but only 
in the data analysis. 
In this way it would be possible the gravitational analogue of the 
discovery of triple star systems with electromagnetic 
timing~\cite{triple}.


\section{The equations of motion}

Let us consider a system composed of three bodies attracting 
one another via gravitational force and assume that their 
typical sizes are small compared with their mutual distances.
Let $\bmit r_i$ and $m_i$ be the radius vector and the mass of the 
{\em i}--th body in a given Galilean reference frame.
In order to simplify the problem the following set of 
variables is introduced:
\begin{eqnarray}
&&\xin = \bmit r_{2}-\bmit r_{1};\qquad 
\xo = \bmit r_{3}-\bmit r_{1};\qquad 
\bmit x_{(cm)} = 
\frac{m_1 \bmit r_1 + m_i \bmit r_2 + m_o \bmit r_3}{m_1 + m_i + m_o}
\nonumber\\
&&m_i = m_2; \qquad m_o = m_3. 
\label{coorel}
\end{eqnarray}
where $\bmit x_{(\lambda)}$ ($\lambda=i,o$) are the relative 
coordinates of the $\lambda$--th body with respect to the first one, 
while $\bmit x_{(cm)}$ are the coordinates of the centre of mass.
The equations therefore become (all through the paper $G=c=1$):
\begin{eqnarray}
\label{new1}
&&\frac{d^{2}\xin}{dt^{2}}{\ +\ }\frac{(m_{1}+m_i)\ \xin}
{\mxi^{3}}{\ }
=\frac{\partial \delta {\cal L}_i}{\partial \xin}\qquad\\
\label{new2}
&&\frac{d^{2}\xo}{dt^{2}}{\ +\ }\frac{(m_{1}+m_o)\ \xo}
{\mxo^{3}}{\ }
=\frac{\partial \delta {\cal L}_o}{\partial \xo}\qquad\\
&&\frac{d^2 \bmit x_{(cm)}}{dt^{2}} = 0
\label{newcm}
\end{eqnarray}
where the so called {\em disturbing functions} have been defined
\begin{eqnarray}
\label{dist1}
&&\delta{\cal L}_{i}(\xin, \xo){\ }={\ }m_o{\ }
\left(\frac{1}{|\xin - \xo|} -
\frac{\xin\cdot\xo}{\mxo^{3}}\right)\\
\label{dist2}
&&\delta{\cal L}_{e}(\xin, \xo)
{\ }={\ }\frac{m_i}{m_o}{\ }\delta{\cal L}_{i}(\xo, \xin)
\end{eqnarray}
We assume that the mutual interaction between $m_i$ and
$m_o$ is a perturbation to their Keplerian motion around 
$m_1$. This is accomplished if the 
following conditions are fulfilled:
\begin{eqnarray}
\mxi &<<& \mxo\\
\frac{m_1+m_i}{m_o} &>>& \left( \frac{\xin}{\xo}\right )^3
\label{maggiore}\\
\frac{m_i}{m_1+m_o} &<<& \left( \frac{\xin}{\xo}\right )^2
\label{minore}
\end{eqnarray}

Within the framework of these assumptions it is possible
to solve the equations of motion by employing
 the so called {\em method of variation of 
parameters} in the form due to Lagrange.
A complete review of this 
method, both in its classical formulation 
and in the post--Newtonian 
extension, 
may be found in Refs.~\cite{cusk,brow,cal97}.

In celestial mechanics the usual way to write  
the solution as a Keplerian motion is~\cite{cusk}:
\begin{eqnarray}
x_\lambda^1&=&\left(\cos(\omega_\lambda)\cos(\Omega_\lambda)-\cos(i_\lambda)
\sin(\Omega_\lambda)\sin(\omega_\lambda)\right)\,\tilde{x}_\lambda^1+
\nonumber\\
&&-\left(\sin(\omega_\lambda)\cos(\Omega_\lambda)+
\cos(i_\lambda)\sin(\Omega_\lambda)\cos(\omega_\lambda)\right)\,
\tilde{x}_\lambda^2\label{xl}\\
x_\lambda^2&=&\left(\cos(\omega_\lambda)\sin(\Omega_\lambda)+\cos(i_\lambda)
\cos(\Omega_\lambda)\sin(\omega_\lambda)\right)\,\tilde{x}_\lambda^1+
\nonumber\\
&&-\left(\sin(\omega_\lambda)\sin(\Omega_\lambda)-\cos(i_\lambda)
\cos(\Omega_\lambda)\cos(\omega_\lambda)\right)\,\tilde{x}_\lambda^2
\label{yl}\\
x_\lambda^3&=&\sin(i_\lambda)\sin(\omega_\lambda)\,\tilde{x}_\lambda^1+
\sin(i_\lambda)\cos(\omega_\lambda)\,\tilde{x}_\lambda^2\label{zl}
\end{eqnarray} 
where
\begin{eqnarray}
&&\tilde{x}_\lambda^1=a_\lambda\,\left(\cos(\eta_\lambda)-e_\lambda
\right)\label{xtl}\\
&&\tilde{x}_\lambda^2=a_\lambda\,\sqrt{1-e_\lambda^2}\,\sin(
\eta_\lambda)\label{ytl}\\
&&M_\lambda = n_\lambda\,(t-T_\lambda) = 
\eta_\lambda-e_\lambda \sin(\eta_\lambda)\label{Ml}\\
&&n_\lambda = \sqrt{\frac{m_1 + m_\lambda}{a_\lambda^3}}\label{nl}
\end{eqnarray}
($\lambda = i, o$).
Solutions are parameterized by means of the 12 Keplerian elements
$a_\lambda$, $e_\lambda$, $M_\lambda$, $w_\lambda$, $\Omega_\lambda$, 
$i_\lambda$~\cite{cusk,brow}. Within the framework of the method of 
variation of the parameters, the solution to the full problem has the 
same form as Eqs.~(\ref{xl})--(\ref{nl}) but with elements slowly 
varying upon time according to Lagrangian planetary 
equations~\cite{cusk,brow}.

\section{The generated gravitational waves}

The formal solution--we have reviewed in the previous Section--allows us 
to achieve the expression of the emitted gravitational wave as a 
function of the orbital elements and time explicitly. Since orbital 
parameters are slowly varying, we neglect their time derivative in
performing the calculation.
First we evaluate the reduced quadrupole 
moment 
of the whole system. According to its usual definition~\cite{ll2}
 we have:
\begin{equation}
D_{\alpha\beta} = \int{\rho \left (3 r^\alpha r^\beta -
r^2 \delta_{\alpha\beta} \right ) d^3\bmit r}
\label{defquadr}
\end{equation}
where, in our case, 
\[
\rho = \sum_{j=1}^{3} {m_j \delta(\bmit r - \bmit r_j)}
\]
Substituting eq.~(\ref{coorel}) into the above relations we get
\begin{equation}
{\cal D}^{\alpha\beta} = {\cal D}^{\alpha\beta}_{(i)} + 
{\cal D}^{\alpha\beta}_{(o)} + {\cal D}^{\alpha\beta}_{(io)} 
\label{quadrupolo}
\end{equation}
where:
\begin{eqnarray}
&&{\cal D}^{\alpha\beta}_{(\lambda)} = \mu_{\lambda}\,\left(
3\,x^{\alpha}_{\lambda}x^{\beta}_{\lambda} - 
\bmit {x^2_{(\lambda)}}\,\delta^{\alpha\beta}\right)
\qquad\lambda = i, o\nonumber\\
&&{\cal D}^{\alpha\beta}_{(io)} = \mu_{io}\,\left[ 
3\,\left(x^{\alpha}_i x^{\beta}_o + 
x^{\alpha}_o x^{\beta}_i\right) - 2\,
\bmit {x_{(i)}}\cdot \bmit {x_{(o)}}\, \delta^{\alpha\beta}\right]
\nonumber\\
&&\mu_{i} = \frac{m_i(m_1 + m_o)}{m_1 + m_i + m_o}\quad
\mu_{o} = \frac{m_o(m_1 + m_i)}{m_1 + m_i + m_o}\quad
\mu_{io} = \frac{m_i m_o}{m_1 + m_i + m_o};\nonumber
\end{eqnarray}
$\alpha$ and $\beta$ run from 1 to 3. 
The first two terms in Eq.~(\ref{quadrupolo}) are the quadrupole 
moments of the two binary systems formed by $m_1$ and $m_i$ (the inner 
one) and by $m_1$ and $m_o$ (the outer one); it retains the usual 
form but orbital elements vary with time.
The third term,--an interaction one--is peculiar of the system as a 
whole.
To clarify these points let us consider the components of 
the quadrupole tensor when eqs.~(\ref{xl}--\ref{ytl}) are taken into 
account. In general one has
\begin{equation}
{\cal D}^{\alpha\beta}_{(\lambda)} = \sum_{i,j} 
d^{\alpha\beta}_{ij}(\omega_\lambda,\Omega_\lambda,i_\lambda)
{\cal Q}^{ij}(\tilde{x}^1_\lambda,\tilde{x}^2_\lambda) 
\label{Dlambda}
\end{equation}
\begin{equation}
{\cal D}^{\alpha\beta}_{(io)} = \sum_{i,j}
d^{\alpha\beta}_{ij}(\omega_i,\Omega_i,i_i,\omega_o,\Omega_o,i_o)
{\cal Q}^{ij}_{io}
(\tilde{x}^1_i,\tilde{x}^2_i,\tilde{x}^1_o,\tilde{x}^2_o)
\label{Dio}
\end{equation}
where $d^{\alpha\beta}_{ij}$ are related to the Euler matrices of the 
rotation linking the intrinsic reference frames of motion to the 
observer reference frame; ${\cal Q}^{ij}$ are the ``intrinsic'' 
quadrupole moment coefficients.
If the orbital elements did not change, the first expression  
would be 
the usual quadrupole tensor of a binary system. The effect of a third 
body thus induces a modulation of this signal. 
As for Eq.~(\ref{Dio}), it is a peculiar feature 
of a three body problem; however the structure of such a term is similar 
to the preceeding ones.

The knowledge of the quadrupole tensor allows us to write the 
TT--gauge components of the gravitational waves emitted by the 
triple system. As it is well known (e.g. 
\cite{ll2,mtw}): 
\begin{eqnarray}
&&h^{TT}_{\ \ ij}{\ =\ }\frac{2}{3|{\bf x}|}{\ }\left[
\left(\frac{d^{2}D_{ij}}{dt^{2}}\right)
-n_{i}n^{k}\left(\frac{d^{2}D_{jk}}{dt^{2}}\right)-
n_{j}n^{k}\left(\frac{d^{2}D_{ik}}{dt^{2}}\right)+ \right.
\nonumber\\
&&\qquad\qquad\left.+\frac{1}{2}\delta_{ij}
n^{l}n^{k}\left(\frac{d^{2}D_{lk}}{dt^{2}}\right)+
\frac{1}{2}n_{i}n_{j}n^{l}n^{k}
\left(\frac{d^{2}D_{lk}}{dt^{2}}\right)\right]
\label{hD}
\end{eqnarray}
where $\bmit n = \bmit r/r$ is the unit vector in the observer 
direction.

In order to obtain analytic expressions for the metric perturbation, 
it is simpler to consider the so called {\em true anomaly} 
$\psi_\lambda$ instead of the eccentric anomaly $\eta_\lambda$. The 
relation between these two angles is given by~\cite{cusk}:
\[
\cos{\eta_\lambda} = \frac{e_\lambda + \cos{\psi_\lambda}}
{1 + e_\lambda \cos{\psi_\lambda}}
\]
Starting from the above relation and the definition of $\eta_\lambda$ 
[see eq.~(\ref{Ml})] it is easy to show that the time derivative 
of $\psi_\lambda$ is given by
\[
\frac{d \psi_\lambda}{d t} =
\frac{n_\lambda (1 + e_\lambda \cos{\psi_\lambda})^2}
{\sqrt{(1-e_\lambda^2)^3}}
\]

In order to put the result in a compact form we introduce the 
Euler Matrices $\bmit{A}_\lambda$ with which 
eqs.~(\ref{xl}--\ref{zl}) can be written as (here $^t$ means 
transposition)
\begin{equation}
\bmit{x}_\lambda = \bmit{A}^t_\lambda \tilde{\bmit{x}}_\lambda
\label{euler}
\end{equation}
The second derivative of the quadrupole tensor 
components can be written as ($\aleph \in \{11,12,0\}$)
\begin{eqnarray}
\ddot{D}^{ij}_{(\lambda)} &=& 3 \mu_\lambda 
\sum_\aleph{\da^{ij}(\omega_\lambda,\Omega_\lambda,i_\lambda)
H^\aleph_\lambda(a_\lambda,e_\lambda,\psi_\lambda)}
\label{ddquadrl}\\
\ddot{D}^{ij}_{(io)} &=& 3 \mu_{io} 
\sum_{r,s=1}^{2}{d_{(io)\,rs}^{ij}(\omega_i,\omega_o,\Omega_i,
\Omega_o,i_i,i_o)
H^{rs}_{(io)}(a_i,e_i,\psi_i,a_o,e_o,\psi_o)}
\label{ddquadrio}
\end{eqnarray}
where 
\begin{eqnarray}
H^{11}_{(\lambda)} &=& 
\frac{d^2\ }{dt^2} 
\left( \tilde{x}^1_\lambda \tilde{x}^1_\lambda \right ) =
 - \frac{m_1 + m_\lambda}{2 a_\lambda 
(1 - e_\lambda^2)} \, \left(
3 e_\lambda \cos{\psi_\lambda(t)} + 4 \cos{2\psi_\lambda(t)} + 
e_\lambda \cos{3\psi_\lambda(t)}\right)\qquad\\
H^{12}_{(\lambda)} &=& 
\frac{d^2\ }{dt^2} 
\left( \tilde{x}^1_\lambda \tilde{x}^2_\lambda \right ) =
- \frac{m_1 + m_\lambda}
{a_\lambda (1 - e_\lambda^2)} \, \left(
3 e_\lambda + 4 \cos{\psi_\lambda(t)} + 
e_\lambda \cos{2\psi_\lambda(t)}\right)\, 
\sin{\psi_\lambda(t)}\\
H^{0}_{(\lambda)} &=& 
\frac{d^2\ }{dt^2} 
\left( \tilde{r}^2_\lambda \right ) =
\frac{2 e_\lambda (m_1 + m_\lambda)}
{a_\lambda (1 - e_\lambda^2)} \, \left(
e_\lambda + \cos{\psi_\lambda(t)}\right)
\end{eqnarray}
\begin{eqnarray}
&& H^{11}_{(io)} = 
\frac{d^2\ }{dt^2} 
\left( \tilde{x}^1_i \tilde{x}^1_o \right ) =
      - {{a_i\,
      \left( 1 - {{e_i}^2} \right) \,
      \left( m_1 + m_o \right) \,\cos (\psi_o(t))\,
      {{\left( 1 + e_o\,\cos (\psi_o(t)) \right) }^2}\,
      \cos (\psi_i(t))}\over 
    {{{a_o}^2}\,{{\left( 1 - {{e_o}^2} \right) }^2}\,
      \left( 1 + e_i\,\cos (\psi_i(t)) \right) }} + \nonumber \\ 
&& - {{a_o\,\left( 1 - {{e_o}^2} \right) \,
      \left( m_1 + m_i \right) \,\cos (\psi_o(t))\,
      \cos (\psi_i(t))\,{{\left( 1 + e_i\,\cos (\psi_i(t))
            \right) }^2}}\over 
    {{{a_i}^2}\,{{\left( 1 - {{e_i}^2} \right) }^2}\,
      \left( 1 + e_o\,\cos (\psi_o(t)) \right) }} + \nonumber \\
&& + {{2\, {\sqrt{m_1 + m_i}}\,
      {\sqrt{m_1 + m_o}}\,\sin (\psi_o(t))\,\sin (\psi_i(t))
      }\over {{\sqrt{a_o\,
          {{\left( 1 - {{e_o}^2} \right) }}}}\,
      {\sqrt{a_i\,{{\left( 1 - {{e_i}^2} \right) }}}}}} 
\label{h11ie}
\end{eqnarray}
\begin{eqnarray}
&& H^{12}_{(io)} = 
\frac{d^2\ }{dt^2} 
\left( \tilde{x}^1_i \tilde{x}^2_o \right ) =
      - {{a_i\,
      \left( 1 - {{e_i}^2} \right) \,
      \left( m_1 + m_o \right) \,
      {{\left( 1 + e_o\,\cos (\psi_o(t)) \right) }^2}\,
      \cos (\psi_i(t))\,\sin (\psi_o(t))}\over 
    {{{a_o}^2}\,{{\left( 1 - {{e_o}^2} \right) }^2}\,
      \left( 1 + e_i\,\cos (\psi_i(t)) \right) }} + 
\nonumber \\
&& - {{a_o\,\left( 1 - {{e_o}^2} \right) \,
      \left( m_1 + m_i \right) \,\cos (\psi_i(t))\,
      {{\left( 1 + e_i\,\cos (\psi_i(t)) \right) }^2}\,
      \sin (\psi_o(t))}\over 
    {{{a_i}^2}\,{{\left( 1 - {{e_i}^2} \right) }^2}\,
      \left( 1 + e_o\,\cos (\psi_o(t)) \right) }} + 
\nonumber \\
&& - {{2\,
      {\sqrt{m_1 + m_i}}\,
      {\sqrt{m_1 + m_o}}\,
      \left( e_o + \cos (\psi_o(t)) \right) \,\sin (\psi_i(t))}
     \over {{\sqrt{a_o\,
          {{\left( 1 - {{e_o}^2} \right) }}}}\,
      {\sqrt{ a_i\,{{\left( 1 - {{e_i}^2} \right) }}}}}}
\label{h12ie}
\end{eqnarray}
\begin{eqnarray}
&&H^{21}_{(io)} = 
\frac{d^2\ }{dt^2} 
\left( \tilde{x}^2_i \tilde{x}^1_o \right ) =
      - {{2\, {\sqrt{{m_1} + {m_i}}}\,
      {\sqrt{{m_1} + {m_o}}}\,
      \left( {e_i} + \cos ({\psi_i}(t)) \right) \,\sin ({\psi_o}(t))}
     \over {{\sqrt{{{{a_o}}}\,
          {{\left( 1 - {{{e_o}}^2} \right) }}}}\,
      {\sqrt{{{{a_i}}}\,{{\left( 1 - {{{e_i}}^2} \right) }}}}}} +
\nonumber\\ 
&&-  {{{a_i}\, \left( 1 - {{{e_i}}^2} \right) \,
      \left( {m_1} + {m_o} \right) \,\cos ({\psi_o}(t))\,
      {{\left( 1 + {e_o}\,\cos ({\psi_o}(t)) \right) }^2}\,
      \sin ({\psi_i}(t))}\over 
    {{{{a_o}}^2}\,{{\left( 1 - {{{e_o}}^2} \right) }^2}\,
      \left( 1 + {e_i}\,\cos ({\psi_i}(t)) \right) }} + 
\nonumber\\
&&-  {{{a_o}\,\left( 1 - {{{e_o}}^2} \right) \,
      \left( {m_1} + {m_i} \right) \,\cos ({\psi_o}(t))\,
      {{\left( 1 + {e_i}\,\cos ({\psi_i}(t)) \right) }^2}\,
      \sin ({\psi_i}(t))}\over 
    {{{{a_i}}^2}\,{{\left( 1 - {{{e_i}}^2} \right) }^2}\,
      \left( 1 + {e_o}\,\cos ({\psi_o}(t)) \right) }}
\end{eqnarray}
\begin{eqnarray}
&&H^{22}_{(io)} = 
\frac{d^2\ }{dt^2} 
\left( \tilde{x}^2_i \tilde{x}^2_o \right ) =
      {{2\, {\sqrt{{m_1} + {m_i}}}\,
      {\sqrt{{m_1} + {m_o}}}\,
      \left( {e_o} + \cos ({\psi_o}(t)) \right) \,
      \left( {e_i} + \cos ({\psi_i}(t)) \right) }\over 
    {{\sqrt{{{{a_o}}}\,{{\left( 1 - {{{e_o}}^2} \right) }}}}\,
      {\sqrt{{{{a_i}}}\,{{\left( 1 - {{{e_i}}^2} \right) }}}}}} + 
\nonumber\\
&&- {{{a_i}\, \left( 1 - {{{e_i}}^2} \right) \,
      \left( {m_1} + {m_o} \right) \,
      {{\left( 1 + {e_o}\,\cos ({\psi_o}(t)) \right) }^2}\,
      \sin ({\psi_o}(t))\,\sin ({\psi_i}(t))}\over 
    {{{{a_o}}^2}\,{{\left( 1 - {{{e_o}}^2} \right) }^2}\,
      \left( 1 + {e_i}\,\cos ({\psi_i}(t)) \right) }} + 
\nonumber\\
&&- {{{a_o}\,\left( 1 - {{{e_o}}^2} \right) \,
      \left( {m_1} + {m_i} \right) \,
      {{\left( 1 + {e_i}\,\cos ({\psi_i}(t)) \right) }^2}\,
      \sin ({\psi_o}(t))\,\sin ({\psi_i}(t))}\over 
    {{{{a_i}}^2}\,{{\left( 1 - {{{e_i}}^2} \right) }^2}\,
      \left( 1 + {e_o}\,\cos ({\psi_o}(t)) \right) }}.
\end{eqnarray}
The above coefficients depend only upon the parameters 
$a_\lambda$, $e_\lambda$ and $M_\lambda$,
while 
\begin{eqnarray}
d_{(\lambda)11}^{ij} &=& 
A_{(\lambda)1}^{\,\ \ \ \ i} A_{(\lambda)1}^{\,\ \ \ \ j} -
A_{(\lambda)2}^{\,\ \ \ \ i} A_{(\lambda)2}^{\,\ \ \ \ j} \\
d_{(\lambda)12}^{ij} &=& 
A_{(\lambda)1}^{\,\ \ \ \ i} A_{(\lambda)2}^{\,\ \ \ \ j} +
A_{(\lambda)2}^{\,\ \ \ \ i} A_{(\lambda)1}^{\,\ \ \ \ j} \\
d_{(\lambda)0}^{ij} &=& 
A_{(\lambda)2}^{\,\ \ \ \ i} A_{(\lambda)2}^{\,\ \ \ \ j} - 
\frac{1}{3} \delta^{ij} \\
\dio{rs}^{ab} &=& 3 \left (\Ai ra \Ao sb + \Ai rb \Ao sa \right )
- 2\ \delta^{ab}\ \sum_{j}{\Ai rj \Ao {sj}{}}
\end{eqnarray}
depend only upon the Euler angles $\omega_\lambda$, $\Omega_\lambda$ 
and $i_\lambda$.
The TT--gauge components of the gravitational field given in
Eq.~(\ref{hD}) can then be written as
\begin{equation}
h_{ij}^{(TT)} = h_{ij}^{(i)} + h_{ij}^{(o)} + h_{ij}^{(io)} 
\label{htt}
\end{equation}
where
\begin{eqnarray}
h_{ij}^{(\lambda)} &=& \sum_\aleph
{\ca^{ij}(\omega_\lambda,\Omega_\lambda,
i_\lambda,\bmit{n})
H^\aleph_\lambda(a_\lambda,e_\lambda,\psi_\lambda)}
\label{hl} \\
h_{ij}^{(io)} &=& \sum_{r,s=1}^{2}
{c^{ij}_{(io)rs}(\omega_i,\omega_o,\Omega_i,\Omega_o
,i_i,i_o,\bmit{n})\, 
H^{rs}_{(io)}(a_i,e_i,\psi_i,a_o,e_o,\psi_o)} \label{hio}
\end{eqnarray}
The values of the coefficients 
$\ca^{ij}(\omega_\lambda,\Omega_\lambda,i_\lambda,\bmit{n})$ and 
$c^{ij}_{(io)rs}(\omega_i,\omega_o,\Omega_i,\Omega_o
,i_i,i_o,\bmit{n})$, connecting the 
Euler angles with the observer--source direction,
are given in appendix.

The signal is therefore the sum of three terms: two binary-like and an 
interaction one. As for the first two terms, they are 
equal in form to the signal emitted by two binary systems~\cite{peters}; 
however the orbital elements are not constants;
this fact results in a modulation of the 
signal. 
The third term in Eq.~(\ref{htt}) is a peculiarity arising from the 
interaction between these two binary systems.
Therefore the signal is described by a Fourier series with
 slowly variable coefficients.
Within the framework of conditions~(\ref{maggiore})--(\ref{minore}) the 
interaction term is of the same order of or greater than the inner 
one. Therefore, when the outer signal can be neglected (this may 
happen, for instance, if the sensitivity of the detector near $n_o$ 
is negligibly small; see following section), the signal form {\em is
different} from that one emitted by a simple two--body system.

\section{Applications}

The time variation of orbital elements, induced by 
the disturbing functions, is very slow and it is quite difficult to 
observe within an interval of time which is of the same order of 
magnitude of the smaller unperturbed period involved. Therefore we are 
interested in the so called ``secular'' variation of the parameters.
For the sake of simplicity we set $i_i = i_o = i$, 
that is to say 
we assume the two orbital planes to coincide. Consequently we have 
$\Omega_i = \Omega_o = \Omega$ (because there is only one line of 
nodes).
As we are only interested in secular variations of the elements, 
the disturbing functions~(\ref{dist1}) and~(\ref{dist2}) may be 
averaged to obtain:
\begin{equation}
\left<\delta{\cal L}_i\right>=\frac{m_o}{a_o}\, \left(1 + 
\frac{a_i^2}{4\, a_o^2}\, \frac{1 + \frac{3}{2}e_i^2}
{\left(1 - e_o^2\right)^{\frac{3}{2}}}\right)
\label{dLimediata}
\end{equation}
\begin{equation}
\left<\delta{\cal L}_o\right>=\frac{m_i}{m_o}\, \left<
\delta{\cal L}_i\right>
\label{dLemediata}
\end{equation}
This result is exact in both $e_i$ and $e_o$; see Ref.~\cite{brow} for 
an expansion up to second order in both eccentricities.
By using Lagrange equations~\cite{cusk,brow}, up to first 
order of approximation, one gets:
\begin{eqnarray}
&&\oti = \nu_i\, t + \otio\\
&&\ote = \nu_o\, t + \oteo\\ 
&&M_i = N_i\, t + \sigma_i\\
&&M_o = N_o\, t + \sigma_o
\end{eqnarray}
where we have put $\oti=\omega_i+\Omega_i,\  
\ote=\omega_o+\Omega_o$. $\otio$, $\oteo$, $\sigma_i$ and $\sigma_o$ are 
constants which are determined once
the initial conditions are given, while
$\nu_\lambda$ are the rates of 
periastron precessions. Their 
values are given by the following relations:
\begin{eqnarray}
&&\nu_i = \frac{3 m_o}
{4 n_i a_o^3}\, \frac{\sqrt{1-e_i^2}}{\sqrt{\left(
1-e_o^2\right)^3}}\label{dni}\\
&&\nu_o = \frac{3 m_i a_i^2}
{4 a_o^5 n_o}\, \frac{\left(1 + \frac{3}{2}e_i^2\right)}
{1-e_o^2}\label{dno}\\
&&N_i = n_i\, \left[1 - \frac{7m_o}
{4 a_o^3 n_i^2}\, \frac{\left(1 + \frac{3}{7}e_i^2\right)}
{\sqrt{\left(1-e_o^2\right)^3}}\right]\label{Ni}\\
&&N_o = n_o\, \left[1 + \frac{2m_i}
{a_o^3 n_o^2} + \frac{3 m_i a_i^2}{4 a_o^5 n_o^2}\, 
\frac{\left(1 + \frac{3}{2}e_i^2\right)}
{\sqrt{\left(1-e_o^2\right)^3}}\right]\label{No}
\end{eqnarray}
By the way, we point out that the averaged Lagrangian given in 
Ref.~\cite{brow}, p. 320, does not yield the correct secular variation 
up to second order in eccentricities. In fact, inspection of 
Lagrangian equations shows that the averaged disturbing function 
should be expanded up to second order in both eccentricities, without 
neglecting terms proportional to $e_i^2\,e_o^2$.

Summing up, the secular effects of the interaction between $m_i$ and 
$m_o$ are a change in the frequency of the unperturbed motion and
appearance of two new frequencies related to the 
periastron precession. 
%

In the framework of assumptions~(\ref{maggiore})--(\ref{minore}),
which are the conditions for the perturbative approach to be allowed, 
we consider two possible systems. The first one is a close binary 
system orbiting around a third body whose mass is 
much greater than the total mass of the other two bodies. In other
 words the order of magnitude of the binary system size 
is much smaller than the mean distance between its centre of mass and 
the third body (the so called ``lunar case'').

The second system is a body orbiting 
around a close binary system whose total mass is much greater than 
that of the far body (the so called ``solar system case'').

\subsection{First case}

Let us consider two stars whose masses are $m_1$ and $m_i<<m_1$ that, 
besides a motion around their centre of mass, are orbiting around a 
third star with a mass $m_o>>m_1+m_i$. This two Keplerian orbits, 
called respectively {\em inner} and {\em outer} orbit, are perturbed 
by the disturbing functions~(\ref{dLimediata})--(\ref{dLemediata}). 
The effect of these perturbative terms, as we have already seen, is to
cause a variation of the elements of the unperturbed Keplerian 
motions.

In order to show the peculiarity of the gravitational signal in such a 
situation we consider the following example in which we have 
assigned: $m_1=2 \cdot 10^{31}\ Kg \approx 10\,m_\odot$, 
$m_i=2 \cdot 10^{29}\ Kg \approx 10^{-1}\ m_\odot$, 
$m_o=2 \cdot 10^{32}\ Kg \approx 100\ m_\odot$, 
$a_i= 10^{9}\ m$, $a_o=10\ a_i$ $|\bmit{R}| = 10\ kpc$, 
$e_i = e_e = 0.25$, $i=0$, $\Omega=0$, $\bmit n = (0,0,1)$, 
$\tilde{\omega}_{i0} = 1.1$, $\tilde{\omega}_{o0} = 0.4$, 
$\sigma_i = 0.3$ and $\sigma_o = 0.1$.
It could be formed by a neutron star close to a black--hole a solar 
diameter apart, under the influence of a farther more massive black--hole.  

The motion of this system could be viewed, 
in zero--th order of approximation, as the Keplerian motion 
of the centre of mass of the two closest stars around the far body 
and the relative motion, Keplerian too, of the binary system. 
The perturbation, due to the interaction among the three bodies, is 
therefore described by the variation of the elements of these two 
Keplerian orbits. The effect results in the appearance of two new 
frequencies (the rates of periastron precession, see Eqs,~(\ref{dni}) 
and (\ref{dno})) in the gravitational 
signal emitted besides the two Keplerian ones (see Eqs~(\ref{nl}),
(\ref{Ni}), and (\ref{No})). 
In this case we find 
\begin{eqnarray}
N_i = 1.14\times 10^{-3}\,\frac{rad}{s};&\qquad&
N_o = 1.21\times 10^{-4}\,\frac{rad}{s};\qquad \nonumber\\
\nu_i = 9.19\times 10^{-6}\,\frac{rad}{s};&\qquad& 
\nu_o = 9.64\times 10^{-10}\,\frac{rad}{s}\nonumber  
\end{eqnarray}   
We notice that a similar effect is due to the 1PN perturbation of the 
inner orbit; in this case the 1PN frequency is given by~\cite{ll2}:
\begin{equation}
\nu_i^{1PN} = \frac{3 (m_1+m_i)}{a_i\,(1 - e_i^2)}\,n_i =
5.56\times 10^{-8}\,rad/s.
\end{equation}
Since our treatment of the phenomena is a classical one, its validity 
holds for observational times smaller than 
$2 \pi/\nu_i^{1PN} \simeq 4\,y$. Therefore, for this kind of 
systems, only frequencies greater than this one, that is to say 
$n_i$, $n_o$ and $\nu_i$ could be appreciate. 

In the previous section we have shown that the gravitational wave 
emitted is the sum of three terms: the first  
(second) one is due to the inner (outer) motion and involves two 
frequencies, that is to say $N_i$ ($N_o$) and 
$\nu_i$ ($\nu_o$); the last one is an interaction term 
which depends on all the four frequencies. From what was said it is 
clear that for the outer term the modulation of signal cannot be 
observed, while, as far as the interaction term is concerned, only two 
frequencies yield an observable modulation.

A gravitational detector presents a sensitivity which depends on the
frequency. If we consider a detector for which the ratio between  
sensitivity near $N_o$ and near $N_i$ is negligible small (such an 
assumption is quite realistic for LISA, see for instance \cite{hough})
then the detectable wave profile is merely the sum of the inner and 
interaction terms even if the greater amplitude is associated with the 
outer motion (as it is in this case).
The output results in a signal with frequency $N_i$ whose 
amplitude is modulated by the two frequencies $N_o$ and $\nu_i$ 
(inner precession rate).

In the figures below (figs. 1--3) we show the gravitational wave 
form for the inner term, the interaction one and their sum. Time is 
given in units of the period of the inner orbit.

\subsection{Second Case}
The second kind of triple--system we consider, involves a 
star of mass $m_1$ and other two bodies (whose masses are $m_i$ and 
$m_o$) moving around it; in addition the following relation holds:
$m_1 \gg m_o \gg m_i$. We assume that the motion of both $m_i$ and 
$m_o$ around the central body $m_1$ is a Keplerian one with orbital 
elements varying with time because of the mutual interaction between 
the two lighter bodies.

Now, as it has already be done in the previous case, we study the 
gravitational waves generated by such a kind of system in a particular 
case. In doing so we have set: 
$m_1 = 2 \times 10^{32} Kg \approx 10^2 m_{\odot}$, 
$m_i = 2 \times 10^{29} Kg \approx 10^{-1} m_{\odot}$, 
$m_o = 2 \times 10^{31} Kg \approx 10 m_{\odot}$,
$a_i= 4 \times 10^{9}\ m$, $a_o=10\ a_i$, 
$|\bmit{R}| = 10\ kpc$
$e_i = e_e = 0.25$, $i=0$, $\Omega=0$, $\bmit n = (0,0,1)$, 
$\tilde{\omega}_{i0} = 1.1$, $\tilde{\omega}_{o0} = 0.4$, 
$\sigma_i = 0.3$ and $\sigma_o = 0.1$. 
In this case the two Keplerian frequencies of 
the inner and outer orbit ($N_i$ and $N_o$ respectively) are: 
\begin{equation}
N_i = 4.57 \times 10^{-4}\,\frac{rad}{s};\qquad\qquad 
N_o = 1.52 \times 10^{-5}\,
\frac{rad}{s}
\end{equation}
while the two periastron precession rates are:
\begin{equation}
\nu_i = 3.65\times 10^{-8}\,\frac{rad}{s};\qquad\qquad
\nu_o = 1.20\times 10^{-10}\,\frac{rad}{s} 
\label{dn}
\end{equation}
The frequency involved in the 1PN perturbation of the inner 
orbit is:
\begin{equation}
\nu_i^{1PN} = \frac{3 (m_1+m_i)}{a_i\,(1 - e_i^2)}\, n_i 
\approx 5.42 \times 10^{-8} \frac{rad}{s}
\label{dnpn}
\end{equation}
From (\ref{dnpn}) it follows that the validity of our 
results holds for times smaller than 
$2 \pi/\nu_i^{1PN} = 1.2 \times 10^8 \,sec \approx 4\,y$, 
therefore we can not appreciate the frequencies 
$\nu_i$ and $\nu_o$. For this reason the detectable 
frequencies of the signal are the only Keplerian ones; 
it follows that the inner and outer term of the wave profile are not 
modulated, while the interaction 
term is modulated by only one frequency.

As in the first case, we show graphically (figs. 4--5) the 
gravitational wave form; the observation time is measured in units 
of the periods of the inner orbit. Also in this case if the ratio 
between the detector sensitivities near $N_o$ and $N_i$ is negligible 
small, then the detectable signal is the sum of the inner and 
interaction terms, which gives a signal with frequency $N_i$ and 
amplitude modulated by the outer frequency $N_o$ (fig. 6).

\section{Conclusions}

In this paper we have studied the gravitational radiation emitted by 
triple--star systems.

We have shown that 
the signal emitted by a triple star system such as that envisaged here 
is the sum of three terms. Two of them are equal in form to the 
signals from two binary systems, each of them modulated by the 
secular variation of the orbital elements.
The last term originates in the three--body 
interaction, not having a two--body counterpart. It 
has a fundamental frequency near $n_i$ (angular frequency of inner 
motion) 
and it is modulated by $n_o$ (angular frequency of outer motion) 
and by the secular variation of orbital elements. If $n_i$ 
is the only frequency in the range of detector sensitivity (for LISA 
such an assumption is quite realistic) then a close binary system in 
interaction with a third body, in the assumptions made through the 
paper, gives rise to an output signal which is different in form from 
the signal belonging to an isolated two--body system.

If the inner motion is in the region of this detector sensitivity, the
waveform could be experimentally extracted by the two time series which 
will be the data of the space based interferometer LISA.
Therefore space based interferometers able to detect 
gravitational radiation from galactic close binary systems will 
make possible 
the discovery and study of eventual three--body systems, without any 
change in their planned design.

In order to test the applicability of this approach we have considered 
two particular cases. We have found that, because of the Newtonian 
approximation we used, there could be systems of interest for which this 
treatment is enough precise to answer for three (first case) or 
two (last case) 
modulating frequencies. 
In this paper we have assumed the
post--newtonian effects involved in motion of the bodies to be 
negligible. Actually, were two of the three bodies very close to each 
other, such effects would become too important to be neglected. 
In this 
case our classical perturbative approach to the motion should be 
replaced by its post--Newtonian extension (see \cite{cal97}, where 
relativistic corrections in a 
simple three--body system are investigated).

\acknowledgments 

The authors are thankful to G. Corbelli, D. Boccaletti, and 
A. Guarnieri for useful discussions, and to V. Guidi for reading of
the manuscript.

\appendix
\section*{}

In the following we give the values of the coefficients 
$\ca^{ij}(\omega_\lambda,\Omega_\lambda,i_\lambda,\bmit{n})$ and 
$c^{ij}_{(io)rs}(\omega_i,\omega_o,\Omega_i,\Omega_o,
i_i,i_o,\bmit{n})$ 
which enter in eqs.~(\ref{hl})--(\ref{hio}). For  
simplicity we have omitted lower indices, being understood that the 
following relations hold for every value of them.


\begin{eqnarray}
c^{11} &=& 
\left (
1-\frac{3}{2} n_1^2 + \half n_1^4 -\half n_2^2-\half n_1^2 n_2^2 
\right)\,d^{11} +
\half \left (
1+n_1^2 \right)\,\left(n_3^2 - n_2^2
\right)\,d^{33} + \nonumber \\
&-& n_1 n_2 \left(1-n_1^2 \right)
\,d^{12} -
n_1 n_3 \left(1-n_1^2 \right)
\,d^{13} +
n_2 n_3 \left(1+n_1^2 \right)
\,d^{23}
\end{eqnarray}


\begin{eqnarray}
c^{33} &=& 
\half \left(1+n_3^2\right) \left(n_1^2 - n_2^2\right)\,
d^{11} +
\left(
1-\frac{3}{2} n_3^2 +\half n_3^4 -\half n_2^2 -\half n_2^2 n_3^2
\right)\,d^{33} + \nonumber \\
&+& n_1 n_2 \left(1+n_3^2\right)\,
d^{12} -
n_1 n_3 \left(1-n_3^2\right)\,
d^{13} -
n_2 n_3 \left(1-n_3^2\right)\,
d^{23}
\end{eqnarray}

\begin{equation}
c^{22} = - c^{11} - c^{33}
\end{equation}

\begin{eqnarray}
c^{12} &=& 
\half n_1 n_2 \left(n_1^2-n_2^2\right)\,
d^{11} +
n_1 n_2 \left(1 - \half n_2^2 + \half n_3^2 \right)\,
d^{33} + \\
&+&\left(1-n_1^2\right) \left(1-n_2^2\right)\,
d^{12} - 
n_2 n_3 \left(1-n_1^2\right)\,
d^{13} -
n_1 n_3 \left(1-n_2^2\right)\,
d^{23}
\nonumber
\end{eqnarray}

\begin{eqnarray}
c^{13} &=&
- n_1 n_3 \left(1-\half n_1^2 + \half n_2^2 \right)\,
d^{11} -
n_1 n_3 \left(1 + \half n_2^2 - \half n_3^2\right)\,
d^{33} + \\
&-&n_2 n_3 \left(1-n_1^2\right)\,
d^{12} +
\left(1-n_1^2\right) \left(1-n_3^2\right)\,
d^{13} -
n_1 n_2 \left(1-n_3^2\right)\,
d^{23}
\nonumber
\end{eqnarray}

\begin{eqnarray}
c^{23} &=&
n_2 n_3 \left(1+\half n_1^2 - \half n_2^2\right)\, 
d^{11} -
\half n_2 n_3 \left(n_2^2 - n_3^2\right)\,
d^{33} + \\
&-& n_1 n_3 \left(1-n_2^2\right)\,
d^{12} -
n_1 n_2 \left(1-n_3^2\right)\,
d^{13} +
\left(1-n_2^2\right) \left(1-n_3^2\right)\,
d^{23}
\nonumber
\end{eqnarray}


\begin{figure}
\caption{Amplitude of the term $h_{11}^{(i)}$ of the signal 
emitted by the triple star system of 
Section 4.1; the orbital plane is perpendicular to the 
observation direction. For the meaning of the symbols used see 
eqs.~(\ref{htt})--~(\ref{hio}). {\em a)} The fundamental frequency is 
$N_i/\pi$. {\em b)} The modulation of the signal has an angular 
frequency given by the rate of periastron precession $\delta n_i$.}
\label{fig1}
\end{figure}

\begin{figure}
\caption{Amplitude of the interaction term $h_{11}^{(io)}$ 
emitted by the triple star system of 
Section 4.1; the orbital plane is perpendicular to the 
observation direction. For the meaning of the symbols used see 
eqs.~(\ref{htt})--~(\ref{hio}). {\em a)} The fundamental frequency is
$N_i/(2\,\pi)$ while the frequency modulation is $N_o/(2\,\pi)$.
{\em b)} The third modulating angular frequency is the periastron 
precession rate $\delta n_i$.}
\end{figure}


\begin{figure}
\caption{Amplitude of $h_{11}^{(io)} + h_{11}^{(i)}$, the 
detectable signal (see considerations in the text) emitted by the 
triple star system of Section 4.1. {\em a)} It is the sum of 
figs. 1a and 2a. {\em b)} It is the sum of figs. 1b and 2b.} 
\end{figure}

\begin{figure}
\caption{Amplitude of the inner term $h_{11}^{(i)}$ of the 
gravitational wave emitted by the triple star system of Section 4.2;
the orbital plane is perpendicular to the observation direction.
For the meaning of the symbols used see 
eqs.~(\ref{htt})--~(\ref{hio}). The fundamental frequency is 
$N_i/\pi$.}
\end{figure}

\begin{figure}
\caption{Amplitude of the interaction term $h_{11}^{(io)}$ 
emitted by the triple star system of 
Section 4.2, when the orbital plane is perpendicular to the 
observation direction. For the meaning of the symbols used see 
eqs.~(\ref{htt})--~(\ref{hio}). The fundamental frequency is 
$N_i/(2\,\pi)$; the modulation is given by $N_o/(2\,\pi)$.} 
\end{figure}

\begin{figure}
\caption{Amplitude of $h_{11}^{(io)} + h_{11}^{(i)}$, 
the detectable signal (see considerations in the text) emitted by the 
triple star system of Section 4.2. It is the sum of figs. 4 and 5.}
\end{figure}

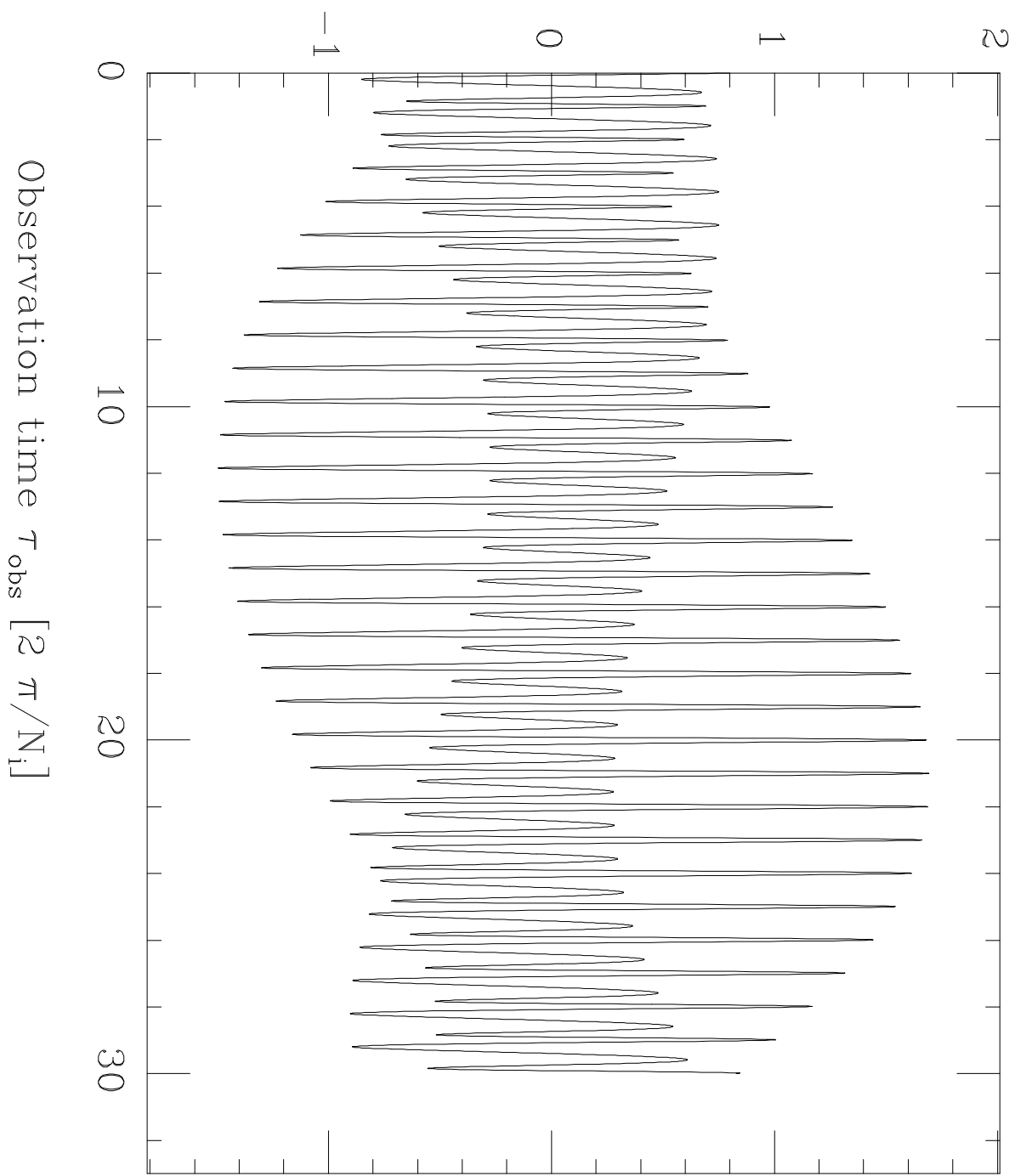
\end{document}